\documentclass[pra,aps,twocolumn,epsfig,showpacs]{revtex4}

\usepackage{graphicx}
\usepackage{rotating}
\usepackage{psfrag}

\begin{document}

\title{Atom optics with rotating Bose-Einstein condensates}

\author{I. Josopait, \L. Dobrek, L. Santos, A. Sanpera 
and M. Lewenstein}

\affiliation{ Institut f\"ur Theoretische Physik, Universit\"at Hannover,
D-30167 Hannover,
Germany}

\begin{abstract}
The atom optics of Bose-Einstein condensates containing a vortex of
circulation one is discussed. We first analyze in detail the 
reflection of such a condensate falling on an atomic mirror. 
In a second part, we consider a rotating condensate in the case of 
attractive interactions. We show that for sufficiently large nonlinearity 
the rotational symmetry of the rotating condensate is broken.
\end{abstract}
\pacs{32.80Pj, 42.50Vk}
\maketitle

\newcommand{\etal}{\mbox{\it et.\ al.\ }}
\newcommand\iso[2]{$^{#2}$#1}
\newcommand\eqref[1]{(\ref{#1})}
\newcommand\myvec[1]{\vec{#1} \hskip 0.5mm}

\newcommand\sequence[7]{
\vbox{
\hbox{
\includegraphics[width=3.85cm]{#1#2.eps}
\hskip 3mm
\includegraphics[width=3.85cm]{#1#3.eps}
}
\vskip 3mm
\hbox{
\includegraphics[width=3.85cm]{#1#4.eps}
\hskip 3mm
\includegraphics[width=3.85cm]{#1#5.eps}
}
\vskip 3mm
\hbox{
\includegraphics[width=3.85cm]{#1#6.eps}
\hskip 3mm
\includegraphics[width=3.85cm]{#1#7.eps}
}
}
}

\newcommand\sequencevier[5]{
\vbox{
\hbox{
\includegraphics[width=3.75cm]{#1#2.eps}
\hskip 5mm
\includegraphics[width=3.75cm]{#1#3.eps}
}
\vskip 5mm
\hbox{
\includegraphics[width=3.75cm]{#1#4.eps}
\hskip 5mm
\includegraphics[width=3.75cm]{#1#5.eps}
}
}
}

\psfrag{mg2Doverh}{${m\over \hbar^2}g_\mathrm{2D}$}
\psfrag{sqrtr2mean}{$\sqrt{<\myvec{r}^2>}$}
\psfrag{sqrtr2emean}{$\sqrt{<(\myvec{r} \cdot \myvec{e_i})^2>}$}
\psfrag{angularmomentumhbar}{\hskip -7mm angular momentum $[\hbar]$}
\psfrag{timeinms}{time [ms]}

\section{Introduction}


During the last years, the experimental achievement of Bose-Einstein 
condensation (BEC) \cite{bec,Hulet} 
has raised a large interest, and numerous 
theoretical and experimental efforts have been devoted to the analysis 
of its intriguing properties. This interest is partially motivated by 
the fact that the BEC is a macroscopic coherent matter wave, with 
evident applications in the already well-developed field of atom optics \cite{Atomoptics}. 
In this sense, the dynamics of BECs interacting 
with atom optical elements, such as e.g. optical or magnetic
mirrors \cite{ertmer,arnold}, or wave guiding \cite{waveguides}, has been recently investigated. 


Additionally, the BEC presents remarkable properties due to its 
superfluidity \cite{Landau}, in particular the quantization 
of vortex circulation \cite{vortexquant}. In this sense, 
studies of vortices in trapped condensates have
attracted a growing attention \cite{jila,ens,mit}. 
A first aim of this paper is the analysis of the atom-optics of a BEC 
that contains a vortex. In particular we study a BEC with a single
vortex of circulation one when falling to and being reflected from an atomic mirror. 
We show that the coherent dynamics exhibits in such a situation
self-interference effects, restoration of a vortex of 
opposite circulation after reflection, splitting of the BEC into two parts at the top of the bounce, 
and formation of additional vortices in low density regions.
This analysis may serve as a way to investigate the properties of 
rotating BECs. 


During the recent years, the development of the Feshbach-resonance 
technique has allowed it to change the strength 
and even the sign of the interparticle interactions \cite{mit_fesh,bosenova}. 
This technique consists of employing magnetic fields to excite resonances 
between atomic and molecular states, and in this way strongly 
influence the value of the $s$-wave scattering length $a_s$ \cite{verhaar}.
By using this novel technique, particularly remarkable experiments have been 
performed, including the Bose-nova experiments at JILA \cite{bosenova}, 
or the recent creation of bright solitons \cite{salomon,huletsoli}.
The same technique can be employed to analyze the stability and eventual 
collapse of a BEC containing one or more vortices. This constitutes 
the second part of this paper. We show that a
larger stability, similar to that discussed for the case of surface modes 
\cite{pattanayak}, can be expected in the case in which the rotational 
symmetry is kept, due to the centrifugal barrier imposed by the vortex. 
However, we show that the rotational symmetry is actually broken when the attractive  
nonlinearity is adiabatically increased to a sufficiently large value. For that case, we show that 
eventually a piecewise collapse is produced for nonlinearities comparable 
with the critical ones for a non rotating condensate.

The paper is organized as follows. In section \ref{sec2} we discuss 
the reflection of a rotating BEC from a mirror. 
In section \ref{sec3} we study the collapse of a rotating BEC, discussing 
the stability and the possibility of breaking of the rotational symmetry.
Finally, in section \ref{sec4} we present some conclusions.

\section{Reflection of a rotating BEC}
\label{sec2}


In the following, we consider a gas of $N$ bosons placed in a harmonic trap 
with pancake symmetry of frequencies $\omega_\perp \ll \omega_z$. 
For sufficiently strong axial confinement, such that $\mu \ll \hbar\omega_z$, 
with $\mu$ the chemical potential, the axial dynamics can be considered 
as effectively frozen. Hence, the wave-function can be written as 
$\psi_\mathrm{3D}(x,y,z)=\psi(x,y)\psi_0(z)$, where the transversal profile 
$\psi_0(z)$ is that of the ground-state of the axial harmonic oscillator. 
For sufficiently low temperatures, the BEC dynamics is then 
well described by the two-dimensional Gross-Pitaevskii 
equation (GPE):
\begin{equation}
\label{GPeq2D}
 i\hbar\partial_t \psi = \left(-{\hbar^2\over 2m}\nabla_{xy}^2 +
 \frac{m}{2}\omega_\perp^2 (x^2+y^2)+ g_\mathrm{2D}|\psi|^2\right)\psi
\end{equation}
with $g_\mathrm{2D} = \sqrt{{m\over \hbar}{\omega_z\over 2\pi}}g_\mathrm{3D}$, 
where $g_\mathrm{3D} = {4\pi \hbar^2 \over m} N a_{s}$, and $m$ the atomic mass.
Demanding by constraint, that the non-interacting BEC has a vortex of
charge $n$ in its center, 
the lowest-energy state in an axially symmetric trap is
\begin{equation}
\label{groundstate}
\psi(r,z,\phi) \propto e^{in\phi}r_\perp^n\exp\left[-{1\over
2}\left(\omega_\perp^2r_\perp^2 + \omega_z^2z^2\right)\right].
\end{equation}

\def\Hz{\hbox{Hz}}


The reflection of a non-rotating BEC has been experimentally accomplished 
both from optical \cite{ertmer} and magnetic \cite{arnold} mirrors.
In the following we analyze such a reflection for the case of a rotating 
condensate.

\begin{figure}[htbp]
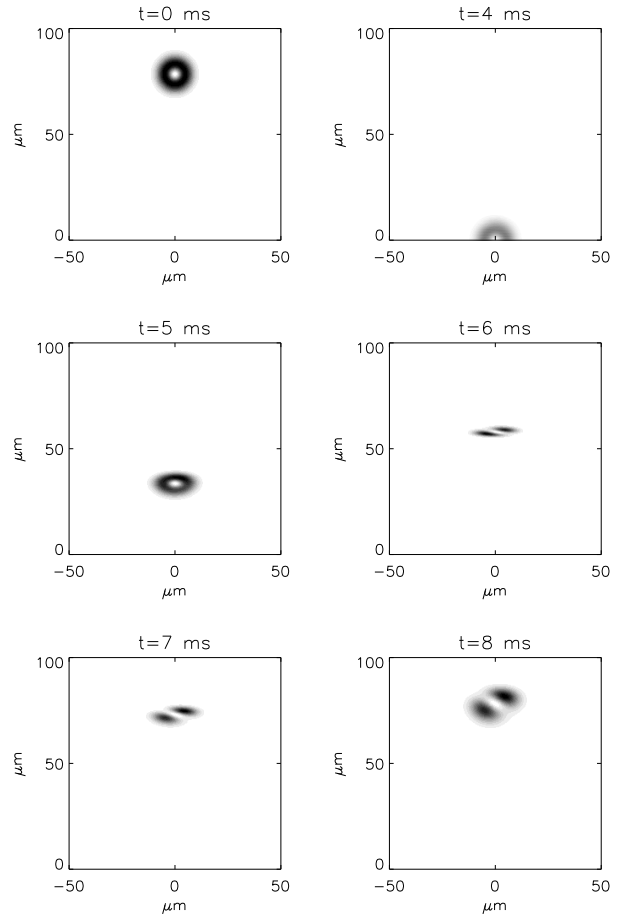
 \begin{center}
\sequence{figures/1/neww2_y_0M_4ms_3000x3000/}{1}{5.2048}{6}{7}{8}{9}
\caption{\label{fig:y_0M} 
Reflection of a rotating BEC for $g_\mathrm{2D}=0$. The gas is released at $t=0$ (top-left figure).
The mirror is placed at the bottom line of the figures.
}
\end{center} \end{figure}

We consider in the following a hard mirror (at $y=0$), 
corresponding to a sufficiently large and abrupt potential barrier. 
In general the mirror has a finite razor, which by means of holographic 
techniques in the case of optical mirrors can be limited to few laser 
wavelengths, typically of the order of few microns. This length scale 
is smaller than those involved in our case (see Figs. \ref{fig:y_0M} 
and \ref{fig:y_10M}), and therefore the softness of the mirror is not 
expected to introduce any significant modification in the discussed effects.
In this case, the mirror can be easily 
simulated by forcing the wave function 
to be antisymmetric on the axis perpendicular to the
mirror: $\psi(y) = -\psi(-y)$. 
Instead of one BEC, which is reflected by a potential, 
we may then consider two condensates:
the ``real'' BEC, and its antisymmetric counterpart on the other 
side of the mirror, the ``ghost'' BEC. 
During the reflection both condensates just
pass through each other, the ghost becoming the observed reflected
BEC.


In our numerical simulations we have set the trap frequencies to
$\omega_\perp = 2\pi \times 20$ Hz and
$\omega_z = 2\pi\times 800$ Hz, which guarantees the two-dimensional 
character of the dynamics for the cases considered. 
By means of imaginary-time evolution \cite{williams} we create
a BEC with a vortex in its center, which is initially located 
at a height $y=h=78.5$ $\mu$m over the mirror. At $t=0$, 
the trap in the $xy$ plane is switched-off ($\omega_\perp=0$). 
Notice that the axial trapping potential remains
switched-on, and therefore the system is maintained two-dimensional. 
The condensate falls
down in the gravity field towards a hard mirror at $y=0$, where it is
bounced upwards again.
We discuss the results for two cases: i) a very low number of atoms 
($g_\mathrm{2D} = 0$), and ii) $N\simeq 18600$ \iso{Na}{23} atoms, 
which corresponds to a coupling parameter 
$g_\mathrm{2D} = 345.6 {\hbar^2\over m}$.
In Figs. \ref{fig:y_0M} and \ref{fig:y_10M} we depict the evolution of 
the falling BEC at different times for the i) and ii) case respectively. 
The condensate is dropped at $t=0$ ms, and the reflection occurs 
at $t\simeq 4$ ms (the mirror is
along the bottom border of the boxes), whereas the upper turning
point is reached at $t \simeq 8$ ms.

In the pictures taken at $t=4$~ms
an interference pattern can be observed of the falling parts of the
condensate and the already reflected parts.
It is not affected by the nonlinearity and shows up
both in linear and nonlinear simulations.
The interference lines are very dense and therefore the pattern is not very 
clearly visible in the figures.
Shortly after $t=6$ ms the BEC is focused to a small strip
by the gravitational cavity. At the top of the bounce the condensates splits into 
two parts (see explanation below) with a region of zero density separating them. 
In the interacting case the core size is smaller, and consequently the 
low-density region which separates both parts is narrower, as observed in 
Fig.\ \ref{fig:y_10M}. In addition, as commented below, in the low-density line we have 
observed in the interacting case the creation of additional vortices \cite{garcia}.

The focusing of the condensate at one point
in its bouncing has classical origins. 
At the reflection gravity changes from an accelerating force to an
decelerating force. Due to the fact that the lower parts of the condensate are reflected
earlier than the upper parts, all atoms gain a small velocity boost towards the center of the BEC.
Indeed, if the velocity spreading $\Delta v$ of the condensate satisfies $\Delta v \ll \sqrt{2gh}$, then one can consider 
the condensate particles initially with $v=0$. In that case if the initial condensate spatial 
width $\Delta y \ll h$, one obtains that after the first bounce the particles are focused at 
$y\simeq 3 h/4$ at 
a time $t=3\sqrt{h/2g}+{\cal O}((\Delta y/h)^2)$. Similar arguments can be applied for larger velocity 
spreadings.

\begin{figure}[htbp] \begin{center}
\sequence{figures/1/later_manymirrorlines_3/}{1}{5.1024}{6}{7}{8}{9}
\caption{
\label{fig:y_10M}
Same as Fig.\ \ref{fig:y_0M} but with $g_\mathrm{2D} = 345.6 {\hbar^2\over m}$.
}
\end{center} \end{figure}

\begin{figure}[htbp]
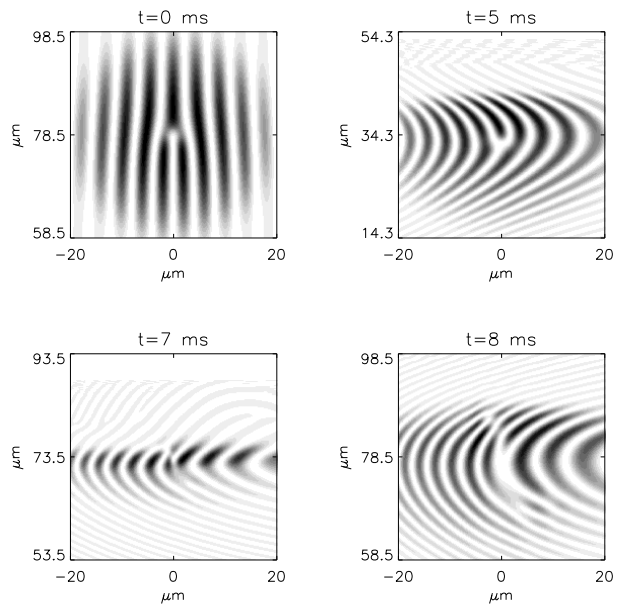
 \begin{center}
\sequencevier{figures/1/l9/}{1.fri}{6.fri}{8.fri}{9.fri}
\caption{
\label{fig:y_10M_fri}
A closer look at the phase of the BEC in Fig.~\ref{fig:y_10M}.
Vortices are visible as fork patterns: Right-spinning vortices show up as forks opened to the bottom, while
left-spinning vortices can be observed as forks opened to the top.
}
\end{center} 
\end{figure}

From the density profiles it is difficult to draw conclusions whether
a vortex remains in the BEC after the reflection. Much more
information can be obtained by looking at the phase of the wave function.
In Fig. \ref{fig:y_10M_fri}, a plane wave is added to the wave function,
$\psi(\vec{r})\rightarrow \psi(\vec{r}) + |\psi(\vec{r})| e^{i\vec{k} \cdot \vec{r}}$.
In this way the phase becomes visible
and a vortex can be detected as a fork in the image \cite{dobrek}. 
In order to be able to observe the phase in the dilute outer regions,
a constant value ($1\over 10$ of
the maximum density) is added to the density.

In the first picture of Fig.\ \ref{fig:y_10M_fri} the initially created (right-spinning) vortex
shows up as a fork which is opened to the bottom.
Shortly after the reflection ($t=5$~ms) the fork is opened to the top, which means that the 
vortex has changed its direction of spin during the reflection and now rotates to the left.
Surprisingly, three vortices are visible in the last picture ($t=8$~ms). Two of them
must have been created during the squeezed state between $t=6$~ms and $t=7$~ms.
Even more interesting is the fact that the total vorticity is not conserved during the squeezing: While at $t=5$~ms
the BEC contains one left-spinning vortex, at $t=8$~ms this vortex is joined by two new
vortices which are both right-spinning.
This vortex instability
does not appear in simulations, in which the nonlinear interactions have been switched off,
and therefore requires nonlinear interactions between the atoms. 
Similar instabilities of vortices have been observed by Garc\'{\i}a-Ripoll {\it et al.} \cite{garcia}
for vortices in trapped condensates.

A rather simple model can be employed to account for the basic features 
of the condensate reflection in the absence of interactions. In that case, the dynamics is provided 
by a Schr\"odinger equation with a potential $V(y)=mg|y|$ (employing the ghost-BEC picture discussed above), 
where $g$ is the gravitational acceleration. It is possible to obtain a 
semi classical approximation by splitting the process into three steps: i) expansion 
while falling towards the mirror, 
ii) reflection, iii) propagation off the mirror.

\def\FT{\hbox{FT}}
\def\rFT{\FT^{-1}}

\begin{figure}[htbp]
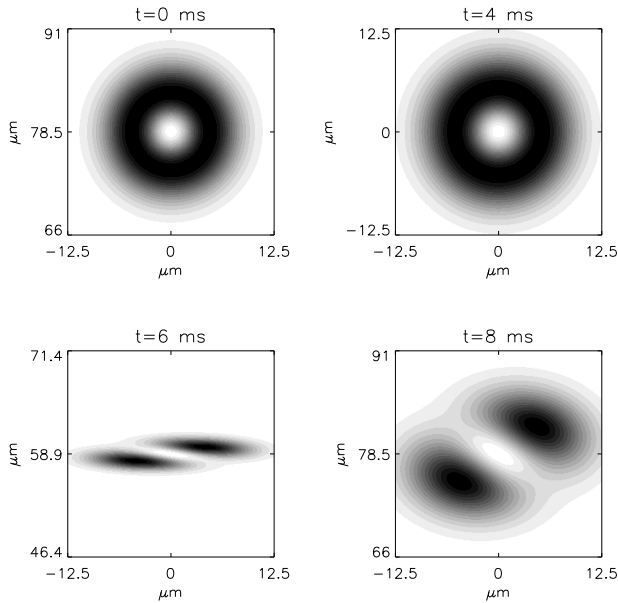
 \begin{center}
\sequencevier{figures/1/propagator2/}{0}{4}{6}{7}
\caption{\label{fig:analytic} 
$\FT$ results for the same case as in Fig.\ \ref{fig:y_0M}.
}
\end{center} \end{figure}

Ignoring the gravitational effects, which will be included later, and evolving from an original wave-function 
$\psi_0(\myvec {x})$, the evolution reduces to a simple expansion in free space:
$\psi_1(\myvec{x},t)=\rFT \left[\exp(-\hbar\myvec{k}^2t/2m) \FT [\psi_0]\right]$,
where $\FT$ and $\rFT$ denote the direct and inverse Fourier transform, respectively.
The main effect of the reflection step ii) is a velocity boost depending on the
position in the condensate. The lower parts of the condensate are
reflected earlier than the upper parts, and this time difference is
essential because at the reflection gravitation changes from an
accelerating force to a decelerating force. We can classically calculate the resulting velocity
boost, $\Delta v(y) = -2y/t_1$, where $t_1=\sqrt{2h/g}$ is the reflection time of the BEC center of mass,
and $y$ is the relative position in the frame of the condensate. 
After applying this velocity boost to the condensate, $\psi_2(\myvec{x}) = \psi_1(\myvec{x},t_1) \exp(-imy^2/\hbar t_1)$, 
we evolve in the stage iii) again in free space until a final time $t_2$ after the reflection of the center of mass: 
$\psi_f(\myvec{x}) = \rFT\left(\exp(-i\hbar\myvec{k}^2t_2/2m)\FT(\psi_2)\right)$.
Fig.\ \ref{fig:analytic} shows the evolution of the BEC wave-function obtained by means of the previously discussed approach, 
which agrees very well with the numerical results shown in Fig.\ \ref{fig:y_0M}. We expect that this 
should be the case as long as the velocity spreading of the condensate $\Delta v \ll \sqrt{2gh}$. 

The previous simplified picture allows for an easy understanding of the breaking up of the 
condensate into two pieces observed in our numerical simulations. In its way up after 
the reflection the condensate possesses a velocity field which results from the addition 
of the vortex velocity field and the linear 
velocity boost resulting from the gravitational field, 
$\vec v=(\omega y/r^2,-\omega x/r^2-2y/t1)$, where $\omega$ is the angular velocity. 
As a consequence the compression is not symmetrical around the vertical coordinate of the 
center of mass, $y=0$, but on the contrary the 
left part of the cloud concentrates in the upper half-plane, whereas the right part does 
it in the lower one. As a result the condensate splits into two pieces.

\section{Implosions of rotating BECs}
\label{sec3}

While in the previous sections the interparticle interactions were assumed to be repulsive, we now consider clouds
of atoms with attractive interactions, i.e. 
$a_s<0$. For small enough trapped condensates, the dispersion induced by the zero-point oscillation of the trap can prevent 
the nonlinear focusing provided by the attractive mean-field, and in this way a metastable BEC 
can be created \cite{Hulet}. However, for sufficiently large condensates in two- and three-dimensional 
trapping geometries, the gas is unstable against collapses. Actually, in physical situations, the formation 
of a singularity is avoided due to the appearance of two- and three-body losses at large densities 
\cite{muryshev,intermittent,santos}. It is the aim of this section to discuss the physics of 
rotating attractive BECs, including stability and the possibility of symmetry breaking.

As a first approach, we consider a vortex at the center of a two-dimensional 
rotationally-symmetric BEC, and assume that the rotational symmetry $ \psi(\myvec{r}, t) =
\chi(r,t) e^{i \phi}$ is kept for any value of the nonlinearity. As we show at the end of this section this 
assumption actually breaks down when the value of $|g_\mathrm{2D}|$ ($g_\mathrm{2D}=-|g_\mathrm{2D}|$) is adiabatically increased to a sufficiently large value.

Our analysis has been performed by means of numerical simulations of the GPE \eqref{GPeq2D}. 
We start with a rotating condensate in the absence of interactions, 
$g_\mathrm{2D}=0$, which is provided 
by \eqref{groundstate} with $n=1$, in the presence of a central vortex.
Then, the coupling parameter $|g_\mathrm{2D}|$ is adiabatically increased
in order to establish the criterion for condensate stability \cite{muryshev}. 
Physically the value of $g_\mathrm{2D}$ can be modified either by increasing the number of atoms for a fixed 
negative value of the scattering length, or by reducing $a_{2D}$ for a fixed number of atoms by means of 
Feshbach resonances. Note that if, on the contrary,  $g_\mathrm{2D}$ is changed rapidly,
additional structures as those discussed in Ref.\ \cite{intermittent} could eventually appear.

We would like to note at this point, that as long as the BEC density is not very large, the GPE \eqref{GPeq2D} 
should describe very well the condensate dynamics. However, as discussed above, at large densities two- and three-body 
losses do play an important role in preventing the formation of a singularity. This could be described by 
including in the GPE the corresponding cubic (two-body) and quintic (three-body) damping terms as discussed 
in Refs.\cite{muryshev,intermittent,santos}. It is not the purpose of this paper to analyze the physics after the collapse 
occurs, and therefore we constrain ourselves to the use of the GPE \eqref{GPeq2D}. 

\begin{figure}[htbp] \begin{center}
\includegraphics[width=8cm]{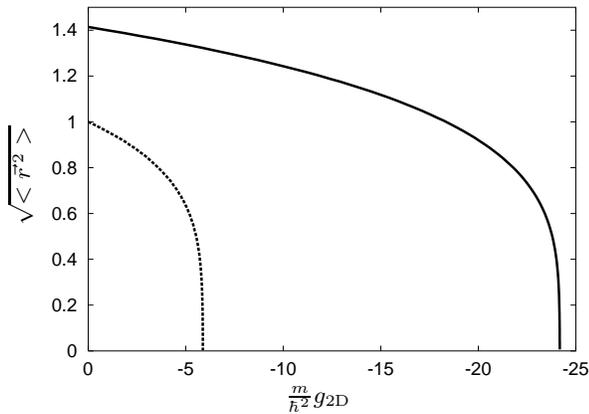}
\caption{
\label{fig:col2d}
Mean square radius $\sqrt{<\myvec{r}^2>}$ for a rotating (solid) and non rotating (dotted) two-dimensional 
BEC if the rotational symmetry is preserved, as a function of $g_\mathrm{2D}$, 
which is adiabatically decreased as linear function of time.
}
\end{center} \end{figure}

In the later stages of the collapse the trapping potential becomes
negligible compared to the mean-field energy and to the kinetic energy. 
Therefore (even for initially non-symmetric confinements) the later stages of the collapse are characterized by 
a single length scale $l(t)$ which has a universal scaling law $l(t) \propto { \sqrt{t-t_0}}$,  
where $t_0$ is the time at which the BEC collapses into a singular point \cite{Zakharov}.
By using this self-similar scaling on the numerical grid it is possible to
perform the simulation up to the very final stages of the collapse.
Fig.\ \ref{fig:col2d} shows the behavior of $\sqrt{\langle r^2 \rangle }$
as a function of the coupling parameter $g_\mathrm{2D}$ for the cases
with and without a vortex at the trap center. 
This condensate width presents the expected scaling at the final stages of the collapse.
As observed in Fig.\ \ref{fig:col2d}, if the rotational symmetry is preserved for any value of $|g_\mathrm{2D}|$
the critical values of the coupling constant, $g_c$, for which the collapse occurs,
significantly differ between the case without and with a vortex. In particular, the 
centrifugal force due to the vortex would amount for a larger stability of the condensate, and consequently for a 
larger value of $|g_c|$. We have numerically obtained that in the presence of vortex the value of $g_c$ is roughly 
four times larger than that in the absence of it: 
$g_c (n=0) = -5.85 {\hbar^2\over m}$ and $g_c(n=1) = -24.15
{\hbar^2\over m}$. A similar stabilization mechanism was discussed
in Ref.\ \cite{pattanayak} for the case of attractive BEC in the presence of surface modes.

An estimation of the critical coupling can be obtained by considering a Gaussian ansatz
\begin{equation}
\psi(r) = {r^n e^{-{1\over 2}{r^2\over \sigma^2}} e^{i\phi n}\over \sigma^{n+1}\sqrt{\pi}},
\end{equation}
and minimizing the corresponding energy functional \cite{dalfovo} 
\begin{equation}
\label{Efunc}
E[\psi] = \int d^2r \left[{\hbar^2\over 2m}|\nabla\psi|^2 +
\frac{m}{2}\omega^2r^2|\psi|^2 + {g_\mathrm{2D}\over 2}|\psi|^4\right].
\end{equation}
In this way we obtain 
$g_c(n=0) = - 2 \pi {\hbar^2\over m} \approx -6.28 {\hbar^2\over m}$,
and 
$g_c(n=1) = - 32 e^{-2} \pi {\hbar^2\over m} \approx -13.6{\hbar^2\over 
m}$.
The Gaussian ansatz provides a good agreement with the numerical results for the case without 
vortex, but underestimates the stability for the case of $n=1$.
This difference may be explained by the departure of the wave-function from the Gaussian ansatz, 
as we have observed in our simulations for the case of coupling constants close to $g_c$.


In our previous numerical simulations, as well as in the analytical estimations, we have assumed that the 
rotational symmetry is preserved when the value of $g_\mathrm{2D}$ is adiabatically changed. 
In the final part of this section we shall show that fully two-dimensional 
simulations clearly indicate that such assumption must be revised, since the 
rotational symmetry is actually broken in the presence of any slight disturbance, when $g_\mathrm{2D}$ is adiabatically increased.

In order to analyze the effects of small perturbations of the rotating BEC, 
we consider a simple model, in which a Gaussian-like condensate with a centered vortex acquires a slight 
asymmetry in the density distribution around the line $x(=r\cos\phi)=0$ 
\begin{equation}
\label{symbreakansatz}
\psi(r,\phi) ={r\,{e^{-1/2\,{\frac {{r}^{2}}{{\sigma}^{2}}}}}{e^{i{\phi}}}
  \left( 1+d\, r
\,\cos \left( {\phi} \right)  \right) 
\over
{\sigma^2\sqrt {
\pi +{d}^{2}\sigma^2\pi }}},
\end{equation}
where $\sigma$ controls the width of the wave function, and $d$ is a small parameter that
controls the asymmetry of the condensate. 
Due to the fact that the sign of $d$ can be chosen arbitrarily, $\partial E/\partial d =0$ at $d=0$, where 
$E$ is defined in Eq.\ (\ref{Efunc}). The second derivative at $d=0$ provides information about the
stability of the condensate:
\begin{equation}
\label{diffd2}
\left. {\partial^2 E\over \partial d^2}\right|_{d=0} =
{1\over 8}\,{\frac {5g_\mathrm{2D} + 8\pi \sigma^4\omega^2}{\pi }}.
\end{equation}
The value for $\sigma$, which minimizes the energy at $d=0$ is given by
\begin{equation}
\left. {\partial E\over \partial \sigma}\right|_{d=0} = {1\over 4}
{8\pi \sigma^4\omega^2 - 8\pi - g_\mathrm{2D}\over \sigma^3\pi} = 0.
\end{equation}
Substituting this expression into Eq.\ \ref{diffd2} we obtain
\begin{equation}
\left. {\partial^2 E\over \partial d^2}\right|_{d=0} =
{1\over 8}\,{\frac{6g_\mathrm{2D}+8\pi}{\pi}}.
\end{equation}
For $0>g_\mathrm{2D}>-{4\over 3}\pi$, this second derivative is positive and the condensate
is expected to be stable.
For $g_\mathrm{2D}<-{4\over 3}\pi$ however, ${\partial^2 \psi\over \partial d^2}|_{d=0}$
is negative. Therefore, it becomes clear even from this very simple model, that 
beyond a given value of $|g_\mathrm{2D}|$ the rotational symmetry breaks down. 

\begin{figure}[htbp] \begin{center}
\sequencevier{figures/2/n29/}{9}{10}{11.3}{23}
\caption{
\label{fig:imprinting_offcentre}
Two-dimensional simulation of a BEC in a vortex state when the value of $g_\mathrm{2D}$ is adiabatically decreased.
The lengths are in units of $\sqrt{\hbar/m\omega}$, and $g_\mathrm{2D}$ in units of $\hbar^2\over m$.
}
\end{center} \end{figure}

\begin{figure}[htbp] \begin{center}
\includegraphics[width=8cm]{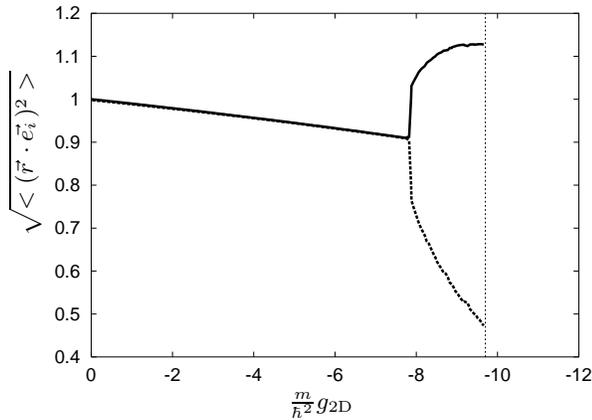}
\caption{
\label{fig:col2d_symbreak2}
The distributions $\sqrt{<(\myvec{r} \cdot \myvec{e}_1)^2>}$ (solid line) and 
$\sqrt{<(\myvec{r} \cdot \myvec{e}_2)^2>}$ (dotted line) of a BEC in a vortex state, 
where $\myvec{e}_1$ and
$\myvec{e}_2$ are the unit vectors pointing into the direction parallel
or perpendicular to the symmetry breaking, respectively, as indicated in 
Fig.~\ref{fig:imprinting_offcentre}. The vertical dotted line indicates the onset of instability.
}
\end{center} \end{figure}

Figure \ref{fig:imprinting_offcentre} shows snapshots taken from 
a two-dimensional simulation. 
A small asymmetry ( corresponding to $d = 0.01 \sqrt{\omega m \over \hbar} $ in Eq.\ (\ref{symbreakansatz}) )
is applied to the initial state. 
Starting at $g_\mathrm{2D}=0$, the coupling parameter is then adiabatically decreased
until the condensate collapses. 
Once a critical value of the coupling parameter is exceeded ($|g_\mathrm{2D}|\simeq 7.8$), 
the asymmetry of the condensate grows and the rotational symmetry breaks. In particular, the density concentrates into two 
oppositely-located peaks. If $|g_\mathrm{2D}|$ is further increased, the width of the peaks decreases and eventually the peaks 
collapse ($|g_c|\simeq 9.7$). Note that the value of $g_c$ is by a factor of about $1.7$ larger than that expected for a non-rotating BEC, 
but since the BEC actually splits into two parts, the critical nonlinearity in each peak is approximately the same as for the 
BEC without vortex. 

The emergence of the symmetry breaking can be observed in more detail in Fig.~\ref{fig:col2d_symbreak2},
which shows the value of  $\sqrt{<(\myvec{r} \cdot \myvec{e}_i)^2>}$ as a function of $g_\mathrm{2D}$, where 
$\myvec{e}_1$ ($\myvec{e}_2$) denotes the direction parallel (perpendicular) to the axis of the symmetry breaking. 
For $g_\mathrm{2D}>-7.8{\hbar^2\over m}$ the rotational symmetry holds and the condensate 
is perfectly described by those calculations which employ such a symmetry. 
For $-7.8{\hbar^2\over m} > g_\mathrm{2D} > -9.7{\hbar^2\over m}$ the rotational symmetry is broken, and 
the condensate splits into two parts orbiting around the trap center. At this stage the larger is $|g_\mathrm{2D}|$ the narrower are the 
peaks of the distribution. Finally, for $g_\mathrm{2D} < -9.7{\hbar^2\over m}$ the condensate is unstable and collapses into two points.

We would like to stress that a very similar picture has been observed due to numerical inaccuracy 
even in the absence of any initially imposed perturbation.  
This strongly suggests that any sort of physical noise, e.g. thermal one, will lead to a breaking of the rotational 
symmetry when the collapse is adiabatically approached.

\section{Conclusions}
\label{sec4}

In this paper we have analyzed some relevant phenomena occurring in the atom-optical physics 
of rotating condensates. In a first part we have analyzed the reflection of a rotating BEC 
from an atomic mirror. We have observed that the vortex is 
preserved after the reflection, both in the presence and in the absence of interactions, 
and that its rotation is inverted. 
The combination of gravitational effects and the mirror produce the squeezing of the BEC cloud 
after its bouncing, and leads eventually to the breaking of the vortex
and the formation of a 
notch in the density. In the interacting case, we have observed the formation of additional vortices 
in the region of the notch.

In the second part of this paper, we have considered the case of a rotating BEC in the presence of attractive interactions. 
We have shown that if the rotational symmetry is conserved, the 
centrifugal barrier induced by the vortex amounts for a larger stability of the system. However, more careful calculations 
show that, if the value of the nonlinearity is adiabatically reduced beginning by the noninteracting case 
(either by increasing the number of condensed atoms or by modifying the value of the scattering length via Feshbach resonances),
the rotational symmetry is actually broken, and eventually a piecewise collapse is produced. Such a symmetry breaking should 
occur for any initial slight noise, i.e. even in the absence of any imposed initial external perturbation. 

We acknowledge support from the Alexander von Humboldt Stiftung, the 
Deutscher Akademischer Austauschdienst (DAAD), the Deutsche 
Forschungsgemeinschaft, the RTN Cold Quantum gases, 
and the ESF Program BEC2000+.

\end{document}